\documentclass[%
 aip,
 sd,%
 amsmath,amssymb,
 reprint,%
]{revtex4-1}

\usepackage{graphicx}
\usepackage{dcolumn}
\usepackage{bm}

\usepackage{amssymb}
\usepackage{color}
\usepackage{amscd}

\usepackage{mathrsfs}
\usepackage{calligra}
\DeclareMathAlphabet{\mathcalligra}{T1}{calligra}{m}{n}

\makeatletter
\newcommand*{\rom}[1]{\expandafter\@slowromancap\romannumeral #1@}
\makeatother

\begin{document}
\preprint{AIP/123-QED}
\title{Interaction between catalytic micro motors}

\author{Parvin Bayati}

\author{Ali Najafi}

 \email{najafi@znu.ac.ir}
\affiliation{Physics Department, University of Zanjan, Zanjan 45371-38791, Iran}%

\date{\today}

\begin{abstract}
Starting from a microscopic model for a  spherically symmetric active Janus particle, 
we study the interactions between two such active motors. 
The ambient fluid mediates a long range hydrodynamic  interaction between two motors. This interaction has both direct and indirect hydrodynamic  contributions. 
The direct contribution is due to the propagation of fluid flow that originated from a moving motor and affects the motion of the other motor. The 
indirect contribution emerges from the re-distribution of the ionic concentrations in the presence of both motors. 
Electric force exerted on the fluid from this 
ionic solution enhances the flow pattern and subsequently changes the motion of both motors. 
By formulating a perturbation method for  very far separated motors, we derive analytic results for the transnational and rotational 
dynamics of the motors. We show that the overall interaction at the  leading order, modifies the 
translational and rotational  speeds of motors which  scale 
as ${\cal O}\left([1/D]^3\right)$ and ${\cal O}\left([1/D]^4\right)$ 
with their separation, respectively.  Our findings open up the way for studying the collective dynamics of 
synthetic micro motors.
\end{abstract}


\date{\today}

\maketitle

\section{Introduction}
Designing the synthetic micro propelling systems with ability to  navigate in predefined 
and controllable trajectories are the aim of many researchers both in chemistry and physics.\cite{molmach}
Delivery of drug  at living systems and construction of manipulating tools for lab on chip experiments 
are among  the main applications of such systems.     
Hydrodynamic swimmers,\cite{laugareview,rojman2s,2s}  single DNA molecule propeller,\cite{dna}
light mediated motion of Janus particles in binary mixtures and colloidal systems\cite{buttinoni,palacci} and
phoretic propulsion of spheroidal particles\cite{popescu} 
are most recent proposed designs for micro machines.

Apart from their potential applications as listed above, the physics of directed motion at the scale of micrometer is also a challenging issue in physics.\cite{brady0,julichercomment} This is mainly due to the inertia-less condition 
that constrains the physics at this scale. At  macroscopic scale of the daily life, inertia provides a 
mechanism for movements, but at microscopic scale, the life is dominated by dissipation. As a result of 
this ambiguity, a backward ejection of a high-speed jet of molecules is not able to propel a micron 
size boat. For driving a micrometer boat, we need to go beyond our macroscopic feeling of 
motion and use nontrivial mechanisms for swimming strategies.\cite{purcell1977life,laugareview,3s}

Janus particles with surface chemical activity are potential proposals for cargo delivery machines 
at the scale of micrometer. 
Originated from a very interesting experiment by Paxton, {\it et. al.},\cite{paxton1,paxton2}
a great deal of the researcher's attention has been attracted  by the idea of generating directed motion by surface reactions.\cite{howse,wang1,wang2,brown}
As a recent example, Janus particles made from spherical Pt insulator have been studied extensively.\cite{ebbens1,ebbens2} 
Drug delivery \cite{baraban,sundararajan,patra,wu,kagan2,mou} and ability for entering into cells   by catalytic Janus particles \cite{gao} 
have been examined experimentally.
Another interesting application of catalytic micro-motors includes water purification that, has been 
successfully tested.\cite{soler,wang4} Also it is shown that using Janus particles, one can make a 
chemical sensor.\cite{kagan1}
Understanding and predicting the physical behavior of a single or many such 
motors constitute the core of many recent researches.\cite{walther}

The physics of a single Janus particle propeller has been investigated theoretically \cite{luga1} and 
numerically.\cite{moran,daghighi,sharifi1} 
Recently, motion of Janus self propellers in different conditions have been considered. This include 
dynamic of a particle confined by  a planar  wall  and also motion in  shear flow. 
It is shown that depending on the initial state of a Janus particle, a rigid and electrically neutral wall can both attract or repel  a 
nearby Janus motor.\cite{uspal,crowdy} The dynamical response 
of a Janus motor moving in an 
ambient shear flow have a crucial dependence on the strength of thermal fluctuations of the ions.\cite{khair}

For most of the expected  applications of micro machines, it is reasonable to use a collection 
of them to achieve maximum efficiency.  
Along this task, one need to have an understanding of the physics of mutual interaction 
between two or many of Janus motors. Such theoretical knowledge will allow the researchers to predict 
the collective behavior  of a suspension of many Janus motors system. So far and up to our knowledge, all theoretical works are 
limited to the physics of individual motors. In this article we address the problem of 
interaction between two self propelled Janus motors. 

A number of interesting phenomena  in a system  
of two or many hydrodynamical swimmers have been observed. These includes a class of phenomena ranging 
from coherent motion of two coupled swimmers \cite{najafi2010coherent,farzin2012general,pooley2007hydrodynamic} to pattern formation and reduction of 
effective viscosity in suspensions.\cite{childress1975pattern,dombrowski2004self,
baskaran2009statistical,hatwalne2004rheology,moradi2015rheological}     
Inspiring from these hydrodynamical systems, we expect to see a rich physical behavior  in 
the case of Janus particles those are electro-hydrodynamical and in addition to 
hydrodynamic effects,  the presence of  
long range electrostatic forces should also be considered.

The rest of this article organizes as follow: In section II we introduce the system and write the basic 
governing equations. Sections III and IV are devoted to develop the approximations that we will use. Analytic results for a single motor are collected in section 
V and the problem of interacting particles is presented in section VI. Finally, discussion about our results 
is presented in section VII.

\section{Governing equations}
\begin{figure}[h!]
 \begin{center}
 \includegraphics[width=0.35\textwidth]{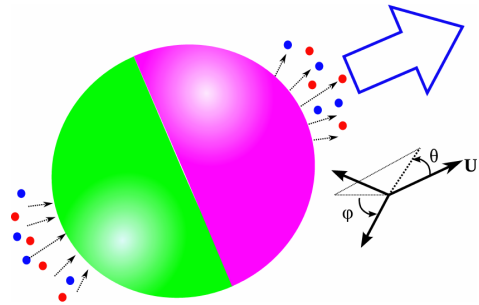}
    \caption{\label{fig1}A spherical micron size particle benefits surface reactions to propel itself. 
    Surface activity of the particle, allow both production and  absorption of ionic molecules.    
    With respect to a co-moving reference frame, an azimuthally symmetric but polar pattern of 
    surface activity, is able to produce a net propulsion.    
   }
  \end{center}
\end{figure}

We start by analyzing the physics of a single motor that, benefits surface chemical reactions to propel itself.
A schematic view of the model system that we are interested to analyze, is shown in Figure~\ref{fig1}. A charged colloidal particle with 
radius $a$ and electrostatic surface potential given by $\psi_s$, is immersed in an electrolyte solution with electric permeability  
$\varepsilon_r$ and hydrodynamic viscosity $\eta$. 
Solution consists of  two ionic species, cations and anions with 
 valances given by $Z_\pm$, respectively (throughout this paper, $+$ refers 
 to cations and $-$ refers to anions). We consider the simple case where the electrolyte is symmetric and single valence 
with $Z_+ = -Z_- =1$.  
Driving force  of this motor  emerges from an asymmetric surface chemical activity. 
We assume that the surface properties of the motor, allows it to absorb and emit chemical species in an asymmetric way. 
As shown in figure, the north hemisphere of the Janus particle 
can emit ionic particles, either 
cations and anions. The south hemisphere of the Janus particle can absorb the ions with the same rate 
given at the emitting part. Such a simple modeling can take into account the physics  of most 
experimentally realized Janus motors.

We  use  dimensionless units to introduce  the dynamical equations. 
In addition to simplifying the notations, dimensionless form of the equations will help us to introduce our approximate 
scheme for  solving the equations. 
We will use the radius of the motor $a$, potential associated with thermal energy $\psi_0=(k_BT/e)$ and equilibrium concentration of ions 
at infinity $n_\infty$,  to make non dimensional form for  all length scales, electric potential and concentrations, respectively.
A velocity scale given by $v_0=(k_BT/e)^2(\varepsilon_r/\eta a)$ and a 
characteristic pressure $p_0=v_0\eta/a$ will be used to make the velocities and pressures non dimensional.

Denoting the ionic fluxes of cations and anions by ${\bf j}_\pm({\bf x})$, we consider a prescribed surface activity given by the following condition on the fluxes:
\begin{equation}
 {\bf \hat{n}} \cdot {\bf j}_{\pm}({\bf r}={\hat {\bf n}}) = {\dot Q} \cos \theta,
\label{surface1}
\end{equation}
where ${\bf j}_\pm({\hat {\bf n}})={\bf j}_{\pm}^{s}$, shows the current densities given on the surface of the spherical Janus particle.  
In the above  equation,  
${\dot Q}$ is a constant ionic rate, $\theta$ is the azimuthal angle with respect to a fixed $z-$axes co-moving with the motor and 
${\bf \hat{n}}$ represents a unit vector that is  normal to the surface. 
For the above model of surface activity, total number  of both ionic species in the solution are fixed.  
In addition to the catalytic realization of our model, it 
can also be considered as a motor that works base on an osmotic pressure difference. 
One can consider a sphere that is constructed by a semi-permeable membrane. An internal active compartment  
inside the motor, provides an angle dependent  osmotic pressure difference between inside and outside of the motor. 
Due to this pressure difference, the above surface ionic flow can appear in the system.

As a result of this asymmetric surface property, the motor will achieve a constant steady state propulsion velocity. 
We would like to calculate this propulsion velocity as a function of the physical properties 
of the motor and  the electrolyte characteristics.

Simultaneous solution to the hydrodynamic equations and the electrostatic equations for the ionic concentrations, will reveal 
the propulsion velocity. These two sets of equations are coupled via the hydrodynamic body force that appears in the hydrodynamic equations. 
As discussed before, the hydrodynamics of a micron scale  system should be described by governing equations at very small inertia condition. 
Neglecting the inertial effects, the Stokes equation governs the dynamics of the fluid at fully dissipative limit:\cite{landau1987fluid}    
\begin{equation}
\nabla ^2 {\bf u}({\bf r})-\nabla p({\bf r})=-(n_+({\bf r})-n_-({\bf r}))\nabla\psi({\bf r}),
\end{equation}
where ${\bf u}({\bf r})$ and $P({\bf r})$ stand for velocity and pressure field of the fluid. Assuming that the fluid is in-compressible, the 
velocity field should satisfy a continuity equation as: $\nabla\cdot\ {\bf u}({\bf r})=0$.
Right hand side of the Stokes equation represents  an electric body force acting on the fluid that comes from the ions, 
where $n_{\pm}({\bf r})$ and $\psi({\bf r})$ are ionic concentrations and electric potential of the ions.  
The electrostatic potential satisfies the Poisson-Boltzmann equation:
\begin{equation}
\delta^2 \nabla^2\psi=-\frac{1}{2}(n_{+} - n_{-}).
\label{Poisson2}
\end{equation}
Continuity equations for the ionic currents are another equations that should be satisfied. The continuity  equations for ions 
 read: 
 \begin{equation}
\frac{\partial n_\pm({\bf r})}{\partial t}+\nabla\cdot{\bf j}_{\pm}({\bf r})=0.
\label{continuty2}
\end{equation}
Thermal fluctuations of the ions, drift due to the electric forces and convection due to the flow of fluid   are 
different sources for the ionic currents.  Collecting all these terms, we can write the following phenomenological relations for 
the ionic currents as:
\begin{equation}
\label{currentdensity}
{\bf j}_{\pm}({\bf r})=-\nabla n_\pm({\bf r})\ \mp  n_\pm({\bf r})\nabla \psi({\bf r})+
{\cal P}e n_{\pm}({\bf r}){\bf u}({\bf r}).
\end{equation}
where the phenomenological contribution from the fluctuations is  expressed in terms of the concentration gradient. 

Two important dimensionless numbers, $\delta$ and ${\cal P}e$ that are appeared in  governing equations, are given by:
\begin{equation}
\delta^2=1/(\kappa a)^2 = \frac{\varepsilon_r k_BT}{2 e^2 n_\infty a^2},~~
{\cal P}e=\frac{\varepsilon_r (k_BT)^2}{\eta e^2 D}.
\end{equation}
Debye screening length $\delta$, measures the equilibrium thickness of ionic cloud around a 
colloid which is immersed in an ionic solution. This is essentially a length beyond which the electric effects of the colloid 
screened.
Peclet number ${\cal P}e$, measures how the 
convection is effective in comparison with thermal diffusion. For very small Peclet number, the current due to the thermal fluctuations is dominated over the current 
from convection. 
In our description of the system, and for mathematical simplifications, the  
diffusion constants for both ions are assumed to be equal and we denote both of them  by $D$. In general the ionic 
diffusion constants  depend  on the size of ions,  that result a different values for cations 
and anions. As we are not interested about the phenomena that could result from such asymmetry between 
cations and anions, we restrict ourselves to the symmetric case with equal diffusion constants. 

In addition to the boundary condition given by equation \ref{surface1}, 
there are other boundary conditions that should be considered. 
On the surface of motor and in a co-moving frame, the boundary conditions for the fluid velocity and ionic potential read:
$$
{\bf u}({\bf r})=0,~~~~\psi({\bf r})=\psi_s, ~~~~~~ {\bf r} =a \hat{{\bf n}}.$$
Very far from the motor, at $r\rightarrow \infty$, the boundary conditions read: 
$$
{\bf u}({\bf r})=-{\bf U},~~~ 
 \nabla \psi({\bf r})=0,~~~~
\psi({\bf r})=0,~~~~
n_\pm= n_\infty,
$$
where ${\bf U}$, is the propulsion velocity of the motor, that needs to be determined by solving the 
above equations. As the motor propulsion is not due to any external force, total force acting on 
the motor vanishes. 
Collecting both the hydrodynamic and electrostatic forces acting on the spherical motor, the force 
free condition can be written as:
\begin{eqnarray}
{\bf F}=\oint_S &&\left\{-p\mathbb{I}+ \nabla {\bf u}+ (\nabla {\bf u})^T\right.\nonumber\\
&&\left. + \nabla \psi \nabla \psi - \frac{1}{2}\nabla \psi\cdot\nabla \psi \mathbb{I}\right\} \cdot{\bf \hat{n}}\ dS=0,
\end{eqnarray}
where $\mathbb{I}$ refers to the $3\times 3$ unit matrix, and 
superscript $T$ refers to transpose of a square matrix.

For a typical experiment that we are interested, 
A micron sized particle with $a\sim 1\mu m$,  moves in an electrolyte solution with $\eta\sim 10^{-3}
\text{Pa~sec}$, $D=\times 10^{-9}\text{m}^2/\text{sec}$  and $n_\infty=10^{23}m^{-3}$ 
(data are given for $0.001$ molar solution of  $KCL$ at room temperature).\cite{ohshima2006theory} 
In this case, we will have $\delta\sim 10^{-3}$ and ${\cal P}e\sim 10^{-1}$. For a typical system with these numerical values for physical parameters, we can 
proceed by applying the condition $\delta\ll 1$  to the dynamical equations and 
obtain approximate analytic results. In the limit of  strong electrostatic screening and also small convection, we expect to 
have great simplifications in dynamical equations.

\section{Thin Debye layer, $\kappa a\gg 1$}
In the limit of thin Debye layer with $\delta\rightarrow 0$, it can be concluded from 
Equation~\ref{Poisson2} that,
\begin{equation}
 n_{+}=n_{-}=N.
\end{equation}
As a result of the singularity at the limit of $\delta=0$, the above electro-neutrality condition can be 
applied only at the outside of Debye layer. The Debye layer, is a thin screening layer adjacent to the surface of Janus particle. 
At this condition and to study the physical properties of the system, we  use a macroscale description developed in 
by Yariv and coworkers.\cite{Yariv1,Yariv2} 
In such description, by decomposing the space into the Debye-layer and the region outside of it (bulk region), 
one  aims to extract an effective macroscale properties of 
the electro-neutral bulk region. 
Such effective fields will provide approximations to the real bulk properties of the system. 
The physics inside the Debye layer is governed by the equilibrium Boltzmann distribution for the 
ionic concentrations. Effective physical properties at the bulk, can be achieved by applying effective
proper boundary conditions on outer surface  of the Debye layer on macroscale fields (instead of the application of boundary conditions on the particle surface).  
Denoting the effective bulk fields by capital letters, we need to solve 
the following governing equations for hydrodynamics in the bulk: 
\begin{equation}
\nabla ^2{\bf V}-\nabla P+\nabla^2 \Psi \nabla \Psi=0,~~~\nabla\cdot\ {\bf V}=0,
\end{equation}
and also the ionic properties are given by the solutions to the following equations:
\begin{equation}
\nabla^2 N-{\cal P}e\ {\bf V}\cdot\nabla N = 0,~~~\nabla\cdot \left(N \nabla \Psi \right)=0.
\end{equation}
The effective fields, obviously satisfy the same boundary  conditions as real microscopic fields at  infinity. 
However, the boundary conditions on the particle surface will change to the boundary conditions given on 
the Debye layer. Combining the boundary condition given in Equation~\ref{surface1} by the definition for the current density 
given by the Equation~\ref{currentdensity}, we arrive at the following boundary conditions on the surface (${\bf r} =\hat{{\bf n}}$):
\begin{equation}
\frac{\partial N}{\partial n}=-{\dot Q}\cos \theta,~~\frac{\partial \Psi}{\partial n}=0,~~
\Psi=\psi_s-\zeta.
\end{equation}
where $\zeta$, is the electric potential drop between the  particle surface and the outer surface of 
Debye layer and $\psi_s$ is the surface potential of the particle. For full  screening 
limit, $\zeta=\psi_s$.
A significant and most important part of the boundary conditions on outer surface of 
Debye layer, is that of a slip velocity
condition that is  known as the 
Dukhin-Derjaguin slip velocity on the effective velocity field given by:\cite{prieve1}
\begin{equation}\label{Dukhin}
{\bf V}_{S}=\zeta \nabla_{S}\Psi - 4\ln\left(\cosh \frac{\zeta}{4}\right)\nabla_{S}\ln N,  ~~~~~~ {\bf r} =\hat{{\bf n}},
\end{equation}
where $\frac{\partial }{\partial n}={\bf \hat{n}}\cdot \nabla $ and 
$\nabla_{S}=(\mathbb{I}-{\bf \hat{n}}{\bf \hat{n}})\cdot\nabla$ is the surface gradient. Here 
$S$  shows the outer surface of Debye layer. 

In the passive case, where the surface of particle is not active, ${\dot Q}=0$ and the above equations have trivial equilibrium solutions given by:
\begin{equation}
N=1,\ \ \ \Psi=0,  \ \ \  P=0,  \ \ \   {\bf V}=0,  \ \ \   \zeta=\text{constant}.
\end{equation}
Obviously for a passive particle, the propulsion  velocity vanishes ${\bf U}=0$. 
We expect to obtain non-zero  self-propulsion velocity for a 
particle with surface activity.

\section{Small surface activity}
Although the thin Debye layer approximation makes the equations very simpler, but they are still 
highly coupled and it is not possible to present analytic solutions.  
This difficulty can be overcome by considering  the case where the surface activity of the 
motor is weak. For very small value of ${\dot Q}$, we can present a systematic expansion in 
powers of ${\dot Q}$. 
Expanding all variables in terms of ${\dot Q}$, we  have:
\begin{eqnarray}
&&N=1+{\dot Q} N'+{\cal O}({\dot Q})^2 , ~~~ \Psi= {\dot Q} \Psi'+{\cal O}({\dot Q})^2,\nonumber\\
&&P={\dot Q} P'+{\cal O}({\dot Q})^2, ~~~~~~~~{\bf V}= {\dot Q} {\bf V'}+{\cal O}({\dot Q})^2,
\end{eqnarray}
and
\begin{eqnarray}
&&\zeta=\zeta_0+{\dot Q} \zeta'+{\cal O}({\dot Q})^2,~~  \psi_s=\zeta_0+{\dot Q} \psi_s'+{\cal O}({\dot Q})^2,\nonumber\\ 
&& {\bf U}={\dot Q} {\bf U}'+{\cal O}({\dot Q})^2.
\end{eqnarray}
Up to the first order in small quantity ${\dot Q}$, the dynamical equations read:
\begin{eqnarray}\label{Stokes}
&&\nabla^2 N'=0 , \qquad  \nabla^2 \Psi'=0 , \nonumber\\
&&\nabla^2 {\bf V'}=\nabla P'  ,  \qquad  \nabla \cdot {\bf V'}=0.
\end{eqnarray}
These equations should be solved provided the following boundary equations at the outer surface of 
Debye layer, ${\bf r} =\hat{{\bf n}}$,
\begin{eqnarray}
&& \Psi'=\psi'_s-\zeta',~~\frac{\partial N'}{\partial n}=-\cos \theta,~~~  \frac{\partial \Psi'}{\partial n}=0 \nonumber\\
&&{\bf V'}=\zeta_0 \nabla_{S}\Psi'-4\ln\left(\cosh \frac{\zeta_0}{4}\right)\nabla_{S}N'.
\end{eqnarray}
The boundary conditions at ${\bf r}\rightarrow \infty$, are given by:
\begin{equation}
N'=0, \qquad \nabla \Psi'=0,  \qquad  {\bf V'}=-U' {\bf \hat{z}}.
\end{equation}
One should note that the force free condition at the first order of ${\dot Q}$ reads:
\begin{equation}
{\bf F'}=\oint_{r=1} \left\{-P'\mathbb{I}+\nabla {\bf V'}+(\nabla {\bf V'})^T\right\}\cdot{\bf \hat{n}}\ dA=0.
\end{equation}
As a result of the above calculations, one can see that in the limit that we work, 
the  electrostatic effects have no contribution in the total force. 

In the following parts, we first derive the propulsion velocity and also the velocity filed due to a 
single motor. Then the problem of interacting motors will be addressed in details.

\section{single Janus particle}
Here, we calculate the properties of a single motor in the limits that described before. 
As $N'$ and $\Psi'$ simply satisfy the Poisson equation,  their 
azimuthal symmetric solutions can be written 
as an expansion in terms of  Legendre polynomials: 
\begin{eqnarray}
\Psi'(r,\theta)&=&\sum_m \left(A_m r^m + B_m r^{-(m+1)}\right)P_m(\cos \theta),\nonumber\\
N'(r,\theta)&=&\sum_m \left(A'_m r^m + B'_m r^{-(m+1)}\right)P_m(\cos \theta).\nonumber\\
\end{eqnarray}
Applying the boundary conditions and up to the leading order of ${\dot Q}$,  
the following unique solutions can be derived:
\begin{equation}\label{N and Psi}
N'=\frac{1}{2 r^2} \cos \theta,~~~~ \Psi'= \psi_s-\zeta=\text{constant}.
\end{equation}
Having in hand the ionic concentration, we can proceed to calculate the hydrodynamical variables as well. 
As a result of symmetry considerations, We  assume that the self propelled velocity of the particle  points along the ${\bf \hat{ z}}$ direction. So  
we can put ${\bf U'}=U' {\bf \hat{z}}$ and search for the value of $U'$.  Using the above result for the concentration profile, the  
slip velocity on the particle surface can be written down as:
\begin{eqnarray}\label{slipcon}
{\bf V'}_{S} &&= -4 \ln \left(\cosh \frac{\zeta_0}{4}\right) \nabla _{S} N'|_{r=1}\nonumber\\
&& = 2 \ln \left(\cosh \frac{\zeta_0}{4}\right) \sin \theta \ {\hat{\pmb \theta}}.
\end{eqnarray}
In addition to the above conditions, the force free condition should also be considered. 

In order to evaluate the particle velocity $U'$, we proceed by applying the well known reciprocal theorem of low Reynolds 
hydrodynamics.\cite{happel2012low}  
The Lorentz reciprocal theorem, relates the solutions of two distinct Stocks flow problems which share the same geometry but 
having different boundary conditions.  According to this theorem, the velocity fields 
${\bf V}_{\rom{1}}$ and ${\bf V}_{\rom{2}}$  and also the corresponding stresses, 
${\pmb \sigma}_{\rom{1}}$ and ${\pmb \sigma}_{\rom{2}}$ of two problems are related by a surface integral over  domain boundaries as:
\begin{equation}\label{Lorentz}
\int {\bf V}_{I} \cdot {\pmb \sigma}_{II} \cdot {\bf \hat{n}}\ dS =\int {\bf V}_{II} \cdot {\pmb \sigma}_{I} \cdot {\bf \hat{n}}\ dS. 
\end{equation}
The  integral is over a surfaces that defines the boundary. 
The velocity profiles, ${\bf V}_{\rom{1}}$ and ${\bf V}_{\rom{2}}$ are subjected to different 
boundary conditions on the surface and both of them are assumed to vanish at infinity. 

\begin{figure}[h!]
 \begin{center}
 \includegraphics[width=0.4\textwidth]{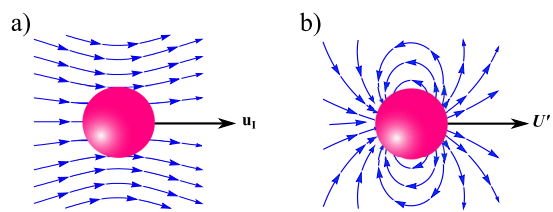}
    \caption{\label{fig2}Two different hydrodynamic problems are used in  reciprocal theorem to achieve the swimming velocity of a Janus particle. a)  Problem \rom{1}: 
    the velocity field of a translating sphere with velocity 
    $u_{\rom{1}} {\bf \hat{ z}}$, b) Problem \rom{2}: the velocity field of a self propelling Janus particle with slip 
    velocity given by Equation~\ref{slipcon}, in laboratory reference frame.}
  \end{center}
\end{figure}

Here and to use the reciprocal theorem for extracting the swimming velocity of Janus particle, we choose the problems I and II as follows. 
For case I, we consider ${\bf V}_{\rom{1}}$ as the velocity field of a translating sphere with an arbitrary velocity given by:  
$u_{\rom{1}}{\bf \hat{ z}}$. This translating sphere is subjected to no slip boundary condition on its surface. 
On the surface of the sphere 
we have: ${\bf V}_{\rom{1}}=u_{\rom{1}}{\bf \hat{ z}}$. As a very well known result, for this translating sphere, the hydrodynamic force has 
a simple form given by: ${\bf F}_{\rom{1}}=-6\pi u_{\rom{1}}{\bf \hat{ z}}$.
For case II, we choose  the velocity profile of our main problem, the problem of a propelling Janus particle with slip velocity. 
We consider the problem of this  
propelling  Janus particle in the laboratory reference frame and put ${\bf V}_{\rom{2}} = {\bf V}' +{\bf U}'$. 
The slip condition on the particle surface is given by:
\begin{eqnarray}
&&{\bf V}_{II}|_{S}={\bf V'}_{S} + {\bf U'},
\end{eqnarray}
where ${\bf V'}_{S}$ is the slip velocity from Equation~\ref{slipcon}. One should note that the hydrodynamic problems of both cases, I and II, vanish at 
infinity.
Streamlines of the above two cases are plotted in Figure~\ref{fig2}. 

After defining the problems I and II, we can easily see that the left-hand-side of Equation~\ref{Lorentz} reads as:
\begin{equation}
\int {\bf V}_{\rom{1}} \cdot {\pmb \sigma}_{\rom{2}} \cdot {\bf \hat{n}}\ dS  =u_{\rom{1}} {\bf \hat{ z}} \cdot \int {\pmb \sigma}_{\rom{2}} \cdot {\bf \hat{n}}\ dS 
=u_{\rom{1}} {\bf \hat{ z}} \cdot {\bf F}_{\rom{2}} =0.
\end{equation}
The last result comes from the force free condition of a self propelling Janus particle. Then from the right-hand-side of Equation~\ref{Lorentz} we have:
\begin{equation}
\int {\bf V}'_{S} \cdot {\pmb \sigma}_{\rom{1}} \cdot {\bf \hat{n}}\ dS = -{\bf U}' \cdot {\bf F}_{\rom{1}} = 6\pi u_{\rom{1}} U',
\end{equation}
where we have used the fact that the force exerted on the particle in our first problem is given by: ${\bf F}_{\rom{1}} = -6\pi u_{\rom{1}}{\bf \hat{ z}}$. 
Substituting Equation~\ref{slipcon} in the above equation, we will have:
\begin{eqnarray}
U'&&=\frac{1}{3\pi u_{\rom{1}}} \ln \left(\cosh \frac{\zeta_0}{4}\right) \oint \sin \theta \ {\hat {\pmb \theta}}\cdot {\pmb \sigma}_{I} \cdot {\bf \hat{n}} dS\ 
\end{eqnarray}
Noting that for a spherical particle, we have ${\bf \hat{n}} \cdot {\pmb \sigma}_{\rom{1}}=-\frac{3}{2} u_{\rom{1}} {\bf \hat{z}}$, the  
propulsion velocity can be obtained as: 
\begin{equation}
{\bf U'}=\frac{4}{3} \ln\left(\cosh \frac{\zeta_0}{4}\right)  {\bf \hat{z}}.
\label{Particle Velocity}
\end{equation}

\begin{figure}[h!]
 \begin{center}
 \includegraphics[width=0.3\textwidth]{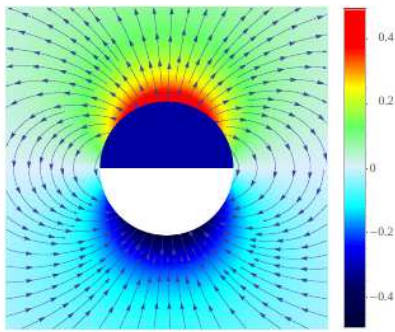}
    \caption{\label{fig3}Ionic density profile and fluid velocity streamlines of an electrokinetic self propelled Janus particle. Asymmetric distribution of ions and 
    slip velocity on the surface are essential elements that cause the Janus particle to move.}
  \end{center}
\end{figure}

In Figure~\ref{fig3}, we have plotted the ionic density profile  and the streamlines of the resulting fluid flow around the self propelled Janus particle. 
Asymmetric distribution of the ions around the Janus particle emerges from the asymmetric surface activity given on the surface of motor. 
Such an asymmetry when combines with the sleep velocity condition given in the macro-scale description, provides an essential physical element for 
producing  a finite self propulsion.  
After returning the physical dimensions, the speed of the Janus particle is given by:
\begin{equation}
U=\frac{4}{3}\frac{\varepsilon_r (k_{B}T/e)^2}{\eta\ a} \frac{{\dot Q} a}{D n_{\infty}} \ln\left(\cosh \frac{e \zeta_0}{4 k_B T}\right).
\end{equation}
The Janus particle moves in a direction that is preferred by the asymmetry of the surface reactions. 
As one can see from the above result, 
both electric effect of Janus particle, that is given by its surface potential $\zeta_{0}$, 
and also the strength of thermal fluctuations $k_{B}T$, 
have dominant influence on the functionality of a single motor. 
In the case of small Peclet number and for large temperature $k_B T\gg e\zeta_0$, 
the case  that we are interested in a typical system, 
the swimming speed can be approximated as: $U\sim (\pi/4)\varepsilon_r{\dot Q}\zeta_{0}^{2}a_{\text{ion}}n_{\infty}^{-1}(k_BT)^{-1}$, where 
we have used the relations $D=k_BT/\xi^{\text{ion}}$ and $\xi^{\text{ion}}=6\pi\eta a_{\text{ion}}$ with the size of ionic molecules 
given by $a_{\text{ion}}$. As one can see, the thermal fluctuations have negative influence on the functionality of Janus motor.
For a typical case with $\varepsilon_r=80\epsilon_0$, ${\dot Q}=10^{7}\mu\text{m}^{-2}s^{-1}$, $a_{\text{ion}}=1\text{nm}$ and $\zeta_0=0.01\text{V}$, 
we arrive at a speed like: 
$U\sim 1 \mu\text{m}~\text{s}^{-1}$. Such speed is completely reasonable to have a functional motor at the scale of micrometer. 

After calculating the propulsion velocity, we can investigate the velocity field due to the self 
propulsion of this single motor. The above approach works only for evaluating the particle speed. In order to 
obtain the velocity field we should solve the Stocks equation with proper boundary conditions. A direct solution to the hydrodynamic equations, presented at 
 appendix
\ref{appendix:a}, reveals that  the velocity field of a self propelled Janus particle in the laboratory frame reads as:
\begin{equation}
\label{Velocity Field-Laboratory}
{\bf V'} = \frac{1}{2} \frac{1}{r^3}U'\left(2 \cos \theta \ {\bf \hat{r}}+ \sin \theta \ {\bf \hat{\theta}}\right) =  
-\frac{1}{2} \frac{1}{r^3} {\bf U'} \cdot \left(\mathbb{I} - 3 {\bf \hat{r}}{\bf \hat{r}}\right).
\end{equation}
One should note that the above velocity field, resembles the velocity field due to a dipole of sink and source of potential flow. 
As a result of the force free condition, we had this expectation from the beginning, that in a multipole expansion of the velocity field, the 
source dipole should have the dominant effect.

\section{Two Interacting Janus particles}
\begin{figure}[h!]
 \begin{center}
 \includegraphics[width=0.4\textwidth]{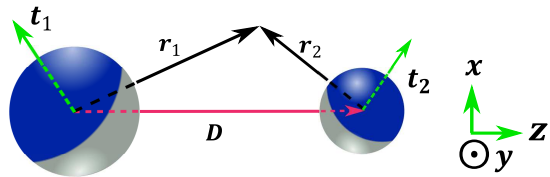}
    \caption{\label{fig4}Geometry of two spherical self propelled Janus particles that are located in distance ${\bf D}$. }
  \end{center}
\end{figure}

As shown in Figure~\ref{fig4},
let us consider two spherically symmetric Janus particles with radii $a_1$ and $a_2$. 
These two particles are separated by a center to center vector denoted by ${\bf D}$. 
The position of a general point in space with respect to each motor is given by ${\bf r}_{1(2)}$, respectively.
Intrinsic propulsion velocities of Janus particles  point along the directions given by ${\bf \hat{t}}_{1(2)}$. In  reference frames that are 
locally connected to each spheres, the 
polar angles are measured with respect to ${\bf \hat{t}}_{1,2}$ and are denoted by 
$\theta_{1,2}$ and $\varphi_{1,2}$. 
The surface  activity of the Janus particles are given by:
$$
{\bf \hat{n}}_{i}\cdot {\bf j}_{\pm}({\bf r}_{i}=a_{i}{\hat {\bf n}}_i)  = {\dot Q}_{i} \cos \theta_{i},~~~i=1,2,
$$
where ${\bf \hat{n}}_{1}$ and ${\bf \hat{n}}_{2}$ represent the unit vectors that are normal to the spheres and the ionic 
production rates at the surfaces of motors are denoted  
 by  ${\dot Q}_{1}$ and ${\dot Q}_{2}$. 
As a result of the calculation given at previous section, the  intrinsic  propulsion speed of  motors, which are the 
speed of isolated motors, are given by Equation~\ref{Particle Velocity}. We want to calculate the influences of a Janus particle on the speed of a 
nearby Janus particle.

At the limit of very small Debye layer, the case that we are interested here, the electric effects 
of particles are screened and the direct electrostatic interaction between the particles can 
be neglected.
In this case, the electro-hydrodynamic forces should be considered. 

Neglecting the direct electrostatic interaction between particles, two types of effects can mediate 
the electro-hydrodynamic forces.  
Both the direct hydrodynamic interaction between moving particles immersed in 
the fluid medium and also the change of fluid pattern due to the rearrangement of ionic species 
which are subjected to instantaneous boundary condition on both particles, 
will eventually lead to coupling between Janus particles. For simplicity, we call the former case by {\it direct} and denote  
the latter case by {\it indirect}  contributions. The direct (indirect) interaction, is due to the 
instantaneous  appearance of particle positions in the boundary conditions of hydrodynamic (electrostatic) equations. 

It is very important to note that both kinds of the above interactions are of hydrodynamics in nature and it is the fluid medium that 
mediates both types of the interactions. 
As an approximation, we assume that these two contributions are additive. The validity of this approximation is guaranteed for very far 
separated particles and we will clarify it in more details at
the following sections.

Denoting the overall translational and rotational velocities of each particles by 
${\bf U}_i$ and $\Omega_i$, we can write them as:
\begin{eqnarray}
{\bf U}_i&=&{\bf U}_{i}^{0}+{\bf U}_{i}^{\text{dir}}+{\bf U}_{i}^{\text{ind}},\nonumber\\
{\bf \Omega}_i&=&{\bf \Omega}_{i}^{0}+{\bf \Omega}_{i}^{\text{dir}}+{\bf \Omega}_{i}^{\text{ind}},
\end{eqnarray}   
where the intrinsic translational and rotational propulsion velocities of the $i'$th Janus particle are 
given by:
\begin{equation}
\label{barevelocity}
{\bf U}^{0}_{i}=\frac{4}{3} {\dot Q}_i\ln\left(\cosh \frac{\zeta_i}{4}\right){\bf {\hat t}}_i,~~~
{\bf \Omega}_{i}^{0}=0.
\end{equation}
From here, we use the radius of the first sphere $a_1$ for making dimensionless lengths. This means that our 
results will depend on a new dimensionless number $e=\left( a_2/a_1\right)$.

We aim to calculate the direct and indirect contributions. 
As described before, the coupling between two particles arises from the boundary conditions that 
 take into account the positions of two particles. 
The ionic concentration field satisfies the Poisson equation and it is 
subjected to boundary conditions 
on the outer part of the Debye  layer of both Janus particles. In the limit of thin 
Debye layer with $\delta\rightarrow 0$, the Poisson equation simplifies to 
${\bf \nabla}^2 N = 0$, where $N({\bf r}_1 , {\bf r}_2)$ shows  the  ionic concentration outside  the 
Debye layer (effective properties of bulk). The ionic concentration satisfies the following boundary conditions:
\begin{equation}
  \frac{\partial N}{\partial r_i} = -{\dot Q}_i \cos \theta_i, ~~~~~~ {\bf r}_i =\hat{{\bf n}}_i, ~~~~i=1,2.
  \label{ionbound}
\end{equation}
In addition to the above boundary conditions, 
the velocity field also should satisfy the Dukhin-Derjaguin slip velocity on the outer surface 
of Debye layer of each particles. The hydrodynamic boundary conditions read:\cite{prieve1}
\begin{equation}
 {\bf V}_{S_i} = -4 \ln \left(\cosh \frac{\zeta_{0_i}}{4} \right) {\bf \nabla}_{S_i} N,~~~ {\bf r}_i =\hat{{\bf n}}_i,~~i=1,2,
 \label{hydbound}
 \end{equation}
where the tangential gradient operator is denoted by ${\bf \nabla}_{S_i}$.

Simultaneous  application of two boundary conditions given by Equation~\ref{ionbound} and Equation~\ref{hydbound}, 
is the main difficulty that makes it impossible to present    a full analytic solution.
In the limit of very far motors, $D\gg 1 , e\sim 1$, we can proceed with a perturbation method. 
We first assume that the motors are hydrodynamically  uncoupled and assume that the  coupling is only due to the 
boundary condition on the ionic concentration.  This will give us the 
indirect contribution. 
Then to obtain an approximation for the direct hydrodynamic contribution, we assume that the motors are 
decoupled with respect to  ionic boundary conditions, 
and investigate the 
hydrodynamic boundary conditions separately. This scheme will provide us a systematic way to expand the 
interaction in powers of $1/D$.  

In the following, we first calculate the indirect contribution and then the direct hydrodynamic contribution is also calculated.

\subsection{Indirect contribution}
To obtain the indirect contribution, we should take into account  the  boundary condition on concentration field that is given by
Equation~\ref{ionbound}. 
For very far particles ($D\gg 1$), we denote the concentration profiles for  isolated particles by:
$N_{1}^{0}$  and $N_{2}^{0}$. In this case we can write the real concentration field that obeys the 
full boundary conditions as:
\begin{equation}
N= N_{1}^{0} + N_{2}^{0}+\Delta N,
\label{pertcon}
\end{equation}
where the deviations from isolated particle solution is denoted by: $\Delta N$.  
As have been calculated in previous sections, the concentration profiles for isolated particles are given by:
\begin{equation}
N_{1}^{0} = \frac{{\dot Q}_1}{2 r_1^2}  \cos \theta_1, ~~ N_{2}^{0} = \frac{e^3 {\dot Q}_2}{2 r_2^2}  \cos \theta_2.
\end{equation}
Writing the deviation from single particle profile as:
\begin{equation}
\Delta N=\sum_{\alpha=1}^{\infty} (N_{1}^{\alpha} + N_{2}^{\alpha})
\end{equation}
we can easily see that the new fields satisfy the Laplace equation as $\nabla ^{2} N_{i}^{\alpha}=0$ with the boundary conditions given by:
\begin{equation}
\frac{\partial N_{i}^{\alpha}}{\partial n}  = -\frac{\partial N_{j}^{\alpha -1}}{\partial n}, ~~~~~ {\bf r}_i =\hat{{\bf n}}_i,~~~~i,j=1,2.
\label{pertboun}
\end{equation}
Such hierarchical description of the effects of motors, allow us to develop a systematic expansion in powers of $1/D$.

At the leading order of calculations, we can proceed by considering the zero order term. 
Having in hand the zero order concentration profile, we can use Equation~\ref{hydbound}
and evaluate the slip velocities on the surface of each motors. Performing such calculations, we can arrive at the following 
relations for the  slip velocities. On the surface of first Janus particle, and in the laboratory reference frame, the velocity reads as: 
\begin{eqnarray}\label{slipind}
&& {\bf V}_{S_1}^{\text{ind}}  = -\frac{3}{2} {\bf U}_1^0 \cdot \left(\mathbb{I} -{\bf \hat{r}}_1 {\bf \hat{r}}_1 \right)\nonumber\\
&& -\frac{3}{2}\frac{e^3}{D^3} \ {\bf U}_2^0 \cdot \left(\mathbb{I} - 3 {\bf \hat{D}} {\bf \hat{D}} \right) \cdot \left(\mathbb{I} -{\bf \hat{r}}_1 {\bf \hat{r}}_1 \right) \nonumber\\
&&+ \frac{9}{2}\frac{e^3}{D^4} \ {\bf U}_2^0 \cdot \left( -2\mathbb{I} + 5 {\bf \hat{D}} {\bf \hat{D}} \right) \cdot  {\bf \hat{r}}_1 {\bf \hat{D}}\cdot \left( \mathbb{I} - {\bf \hat{r}}_1 {\bf \hat{r}}_1\right)\nonumber\\
&&-\frac{9}{2}\frac{e^3}{D^4}\ {\bf \hat{r}}_1 \times ({\bf U}_2^0\times {\bf \hat{D}}),
\end{eqnarray}
and on the surface of second Janus particle, the slip velocity reads as:
\begin{eqnarray}
&&{\bf V}_{S_2}^{\text{ind}} = - \frac{3}{2}\frac{1}{e^3} \ {\bf U}_2^0 \cdot \left(\mathbb{I} - {\bf \hat{r}}_2 {\bf \hat{r}}_2 \right)  \nonumber\\
&& - \frac{3}{2}\frac{1}{D^3} \ {\bf U}_1^0 \cdot \left(\mathbb{I} - 3 {\bf \hat{D}} {\bf \hat{D}} \right) \cdot \left(\mathbb{I} - {\bf \hat{r}}_2 {\bf \hat{r}}_2 \right)  \nonumber\\
&&- \frac{9}{2}\frac{e}{D^4}\  {\bf U}_1^0 \cdot \left( -2\mathbb{I} + 5 {\bf \hat{D}} {\bf \hat{D}} \right) \cdot  {\bf \hat{r}}_2 {\bf \hat{D}}\cdot \left( \mathbb{I} - {\bf \hat{r}}_2 {\bf \hat{r}}_2\right)\nonumber\\
&&+\frac{9}{2}\frac{e}{D^4}\ {\bf \hat{r}}_2 \times ({\bf U_1^0}\times {\bf \hat{D}}).
\end{eqnarray}
As one can see, the slip velocity on each particle has two contributions. As an  example and, for the first particle, these 
two contributions are 
the term  
that is proportional to ${\bf U}_1^0$ and the terms that are proportional to ${\bf U}_2^0$. The first term denotes the 
intrinsic asymmetry of the particle while, the other parts are due to the asymmetry of the second Janus particle. 

As a result of the above surface slip velocities, the velocity filed in the medium  deviates from its value for  the isolated 
Janus particles.
Now to obtain the full changes in the  velocity field, we need to apply  the full hydrodynamic boundary conditions  
on both particle. Here, we want to neglect such complexities and simply assume that the velocity profile is still 
due to 
the isolated Janus particles but with modified slip velocities given by the above relations. 
This is the core of our direct-indirect separation of the effects and what we obtain with this assumption, 
contains the indirect contribution. At the next section, we will again come back to this point and take into account the  
effects that we  neglected here.

We write the fluid velocity field in the laboratory frame as:
\begin{equation}
{\bf V}^{\text{ind}}={\bf V}_{1}^{\text{ind}}+{\bf V}_{2}^{\text{ind}},
\end{equation} 
where the partial  flows  due to each Janus particles can be written as:
\begin{eqnarray}\label{Kinetic-Int-Fluid-Velocity}
 {\bf V}_{1}^{\text{ind}}  =-\frac{1}{2}\frac{1}{r_1^3}\left({\bf U}_{1}^{0} + {\bf U}_{1}^{\text{ind}} \right) \cdot 
 \left(\mathbb{I} - 3 {\bf \hat{r}}_1 {\bf \hat{r}}_1 \right)+{\cal O}(\frac{1}{r_1})^4,\nonumber\\
 {\bf V}_{2}^{\text{ind}} = -\frac{1}{2}\frac{e^3}{r_2^3}\left({\bf U}_{2}^{0} + {\bf U}_{2}^{\text{ind}}\right)\cdot 
 \left(\mathbb{I} - 3 {\bf \hat{r}}_2 {\bf \hat{r}}_2 \right)+{\cal O}(\frac{1}{r_2})^4,\nonumber\\
\end{eqnarray}
where the  velocity contributions that the particles achieved from indirect interaction are denoted by: 
${\bf U}_{i}^{\text{ind}}$ and ${\bf \Omega}_{i}^{\text{ind}}$. The above relations are essentially the velocity profiles of 
isolated Janus particles given in equation \ref{Velocity Field-Laboratory}, 
but with modified swimming velocities. 

Two important conditions  of zero total force and zero total torque, are essential points that we should apply to the 
equations for obtaining ${\bf U}_{i}^{\text{ind}}$ and ${\bf \Omega}_{i}^{\text{ind}}$. 
Similar to the case of a single Janus particle, we can  use the Lorentz reciprocal theorem and extract the 
required results.  
The details of such calculations are collected in appendix~\ref{appendix:b}. For the 
first particle, the final results read:
\begin{eqnarray}
{\bf U}_{1}^{\text{ind}}&=& \frac{e^3}{D^3}U_2^0\ {\bf \hat{t}}_2 \cdot \left(\mathbb{I}-3{\bf \hat{D}}{\bf \hat{D}}\right)\nonumber\\
{\bf \Omega}_{1}^{\text{ind}}&=&-\frac{9}{2} \frac{e^3}{D^4} U_2^0 \left( {\bf \hat{t}}_2\times{\bf \hat{D}}\right),
\end{eqnarray}
and for the second particle, we will have:
\begin{eqnarray}
{\bf U}_{2}^{\text{ind}}&=&\frac{1}{D^3}U_1^0\ {\bf \hat{t}}_1 \cdot \left(\mathbb{I}-3{\bf \hat{D}}{\bf \hat{D}}\right)\nonumber\\
{\bf \Omega}_{2}^{\text{ind}}&=&\frac{9}{2} \frac{e}{D^4} U_1^0 \left( {\bf \hat{t}}_1\times{\bf \hat{D}}\right).
\end{eqnarray}
The above results, present the zero order 
indirect contributions, regarding the perturbation expansion introduced in equation~\ref{pertcon}. 
As one can see, the results for transnational velocity decays like $(1/D)^3$. 
To see the effects of the 
next order terms, we need to solve the Laplace equation for $N_{i}^{1}$ with proper boundary condition given 
by Equation~\ref{pertboun}. As the boundary condition decays like $(1/D)^2$ and the governing 
equation is also linear, we will 
expect such a similar decay for $N_{i}^{1}$. Now for calculating the effects of such first order term in the 
velocities, we should repeat the same procedure as described for zero order term. 
Continuing the calculations, we will eventually receive at a velocity correction for Janus particles that 
decays like $(1/D)^3\times(1/D)^2$. 
At the next part, we will show that the direct hydrodynamic 
contribution give corrections that are more effective than these corrections. So we will ignore the higher order 
corrections due to the concentration profile and keep only the zero order contribution.

\subsection{Direct hydrodynamic contribution}
In the previous section, we have neglected the complexities associated with simultaneous applications of the
slip velocity condition on both particles. We have simply assumed that the velocity profile corresponds to 
the isolated Janus particles but with modified slip velocities. 
Here, we want to go beyond this simplifications and obtain the corrections associated to such complexities. 
To achieve the overall velocity field ${\bf V}$, associated to the 
complete problem that takes into account both direct and indirect interactions, a Stokes equation with the 
following boundary conditions on the surface of Janus particles should be solved:
\begin{eqnarray}\label{Velocity Boundary Cond.-dir}
&&{\bf V}|_{S_1} = {\bf U}_1^0 + {\bf U}_{1}^{\text{ind}} +{\bf \Omega}_{1}^{\text{ind}}\times {\bf r}_1+ {\bf U}_{1}^{\text{dir}} +{\bf \Omega}_{1}^{\text{dir}}\times {\bf r}_1+ {\bf V}_{S_1}^{\text{ind}} \nonumber\\
&&{\bf V}|_{S_2} = {\bf U}_2^0 + {\bf U}_{2}^{\text{ind}} +{\bf \Omega}_{2}^{\text{ind}}\times ({\bf r}_1-{\bf D}) \nonumber\\
&&~~~~~~~~~+ {\bf U}_{2}^{\text{dir}} +{\bf \Omega}_{2}^{\text{dir}}\times ({\bf r}_1-{\bf D})+ {\bf V}_{S_2}^{\text{ind}}\nonumber\\
&& {\bf V}|_\infty  = 0,
\end{eqnarray}
where we have assumed that as  a result of hydrodynamic interaction between the particles, each Janus particle achieve an 
additional change in its velocities given by:  ${\bf U}_{i}^{\text{dir}}$ and ${\bf \Omega}_{i}^{\text{dir}}$. 
Again a proper application of Lorentz reciprocal theorem, will help us to extract the required velocities.  
The details of calculations are presented in  appendix~\ref{appendix:c}, and here we write the final results. For the 
first Janus particle and up to the order ${\cal O}(1/D)^6$, we will have: 
\begin{eqnarray}
{\bf U}_{1}^{\text{dir}} && = - \frac{1}{2}\frac{e^3}{D^3} U_2^0\ {\bf \hat{t}}_2\cdot \left(\mathbb{I} - 3 {\bf \hat{D}} {\bf \hat{D}} \right),\nonumber\\
&& - \frac{1}{2} \frac{e^3}{D^6} U_1^0\ {\bf \hat{t}}_1 \cdot \left(\mathbb{I}-3{\bf \hat{D}}{\bf \hat{D}}\right) \cdot \left(\mathbb{I}-3{\bf \hat{D}}{\bf \hat{D}}\right),
\end{eqnarray}
the first term that is proportional to $U_2^0$, is the velocity field produced by the second Janus particle and calculated at the 
position of first Janus particle. This is a result that we  expected to see from the Faxen theorem for a colloidal particle 
immersed in external velocity field. 
Calculations show that, the dominant part of the rotational velocity induced by direct interaction, will behave like: $(1/D)^9$ and, we neglect it here. 
Similar expression can be obtained for the second Janus particle that reads as:
\begin{eqnarray}
{\bf U}_{2}^{\text{dir}} && = - \frac{1}{2}\frac{1}{D^3} U_1^0\ {\bf \hat{t}}_1\cdot \left(\mathbb{I} - 3 {\bf \hat{D}} {\bf \hat{D}} \right),\nonumber\\
&& - \frac{1}{2} \frac{1}{D^6} U_2^0\ {\bf \hat{t}}_2 \cdot \left(\mathbb{I}-3{\bf \hat{D}}{\bf \hat{D}}\right) \cdot \left(\mathbb{I}-3{\bf \hat{D}}{\bf \hat{D}}\right).
\end{eqnarray}
Very interestingly, the dominant part of both direct and indirect contributions behave similarly for $D\gg 1$. 
Therefore the dominant part of the total translational and rotational velocity of the first Janus particle is given by:
\begin{eqnarray}
{\bf U}_{1} && ={\bf U}_1^0 + \frac{1}{2}\frac{e^3}{D^3} U_2^0\ {\bf \hat{t}}_2\cdot \left(\mathbb{I} - 3 {\bf \hat{D}} {\bf \hat{D}} \right),\nonumber\\
{\bf \Omega}_{1}&&=-\frac{9}{2} \frac{e^3}{D^4} U_2^0 \left( {\bf \hat{t}}_2\times{\bf \hat{D}}\right),
\end{eqnarray}
As a result of symmetry, the corresponding velocities of the  second Janus particle can be written as:
\begin{eqnarray}
{\bf U}_{2} && ={\bf U}_2^0 + \frac{1}{2}\frac{1}{D^3} U_1^0\ {\bf \hat{t}}_1\cdot \left(\mathbb{I} - 3 {\bf \hat{D}} {\bf \hat{D}} \right),\nonumber\\
{\bf \Omega}_{2}&&=\frac{9}{2} \frac{e}{D^4} U_1^0 \left( {\bf \hat{t}}_1\times{\bf \hat{D}}\right).
\end{eqnarray}
In the next part, we show how such interactions will modify the trajectories of Janus particles.
\section{Results and discussion}
\begin{figure}
 \begin{center}
 \includegraphics[width=0.40\textwidth]{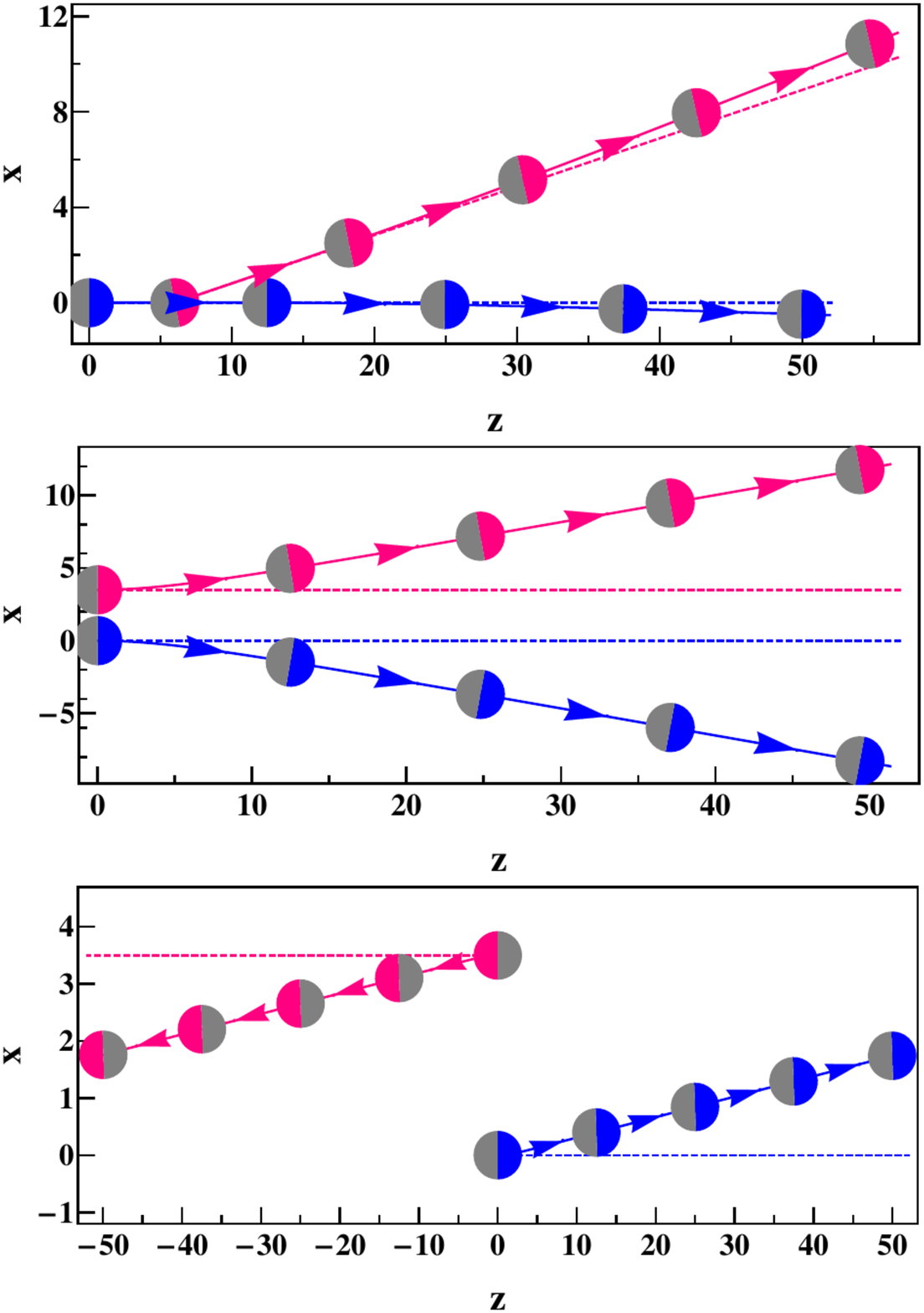}
    \caption{\label{fig5}Trajectories of two interacting Janus particles are shown for different initial 
    states of the motors. In all figures the intrinsic speed of the motors are assumed to be similar. 
    As one can see, for the case of two parallel motors, the interaction tends to repel the Janus 
    particles. }
  \end{center}
\end{figure}

We have shown that as a result of coupling between hydrodynamic and electrostatic effects, a long-range interaction between 
particles have been mediated. 
For a very thin Debye layer $\kappa a\gg 1$ and $a\ll D$, the 
electrostatic effects of Janus particles are screened and we have neglected the  
electric interaction of the Janus particles.  In this case all the interactions between particles have 
hydrodynamical origin that is long range. 
Such long range interaction, affects the translational velocity of each 
motors, and it also introduces a rotational velocity for motors. For very far Janus particles, the 
leading order of translational speed scales 
as ${\cal O}\left([1/D]^3\right)$ and angular velocity scales as 
${\cal O}\left([1/D]^4\right)$.  The interaction  modifies the trajectories of self propelling 
Janus particles.   To have  a qualitative feeling of the interaction, some typical examples of the 
trajectories are presented in Figure~\ref{fig5}. As one can distinguish from figures, the overall effect 
of the interaction depends on the initial states of two Janus particles.  In all trajectories, the dashed-line 
corresponds to the trajectory of an isolated Janus particle. For the cases where the particles move in same directions (the first two 
trajectories shown in figure), the interaction has a repulsive signature. But for a case where the particles move in anti-parallel  
directions (the third trajectory shown in Figure~\ref{fig5}), the interaction tends to decrease the relative speed of two particles.
We have used the dominant part of the interactions but please note that by using the 
method that we have developed in this paper, we are able to systematically consider all of the orders of perturbation terms. 

Along this work, we are currently working on the dynamics of a collection of active Janus  
particles. It is a known fact that hydrodynamic interaction near a rough wall will introduce nontrivial 
effects,\cite{rad2010hydrodynamic} in this case we are also analyzing the motion of a single 
Janus motor near a wall with surface roughness. 

\appendix

\section{Velocity field due to a single  active Janus particle}
\label{appendix:a}
In this appendix, we present the detail of the calculations for the velocity field produced  by a moving single 
Janus motor. 
Along this task, we should solve the Stocks equation, with boundary conditions given as follows:
\begin{eqnarray}
{\bf V'}|_\infty &&= -U' {\bf \hat{ z}} = - U' \cos \theta \ {\bf \hat{r}}+ U' \sin \theta \ {\hat {\pmb \theta}},\nonumber\\
{\bf V'}|_{r=1} &&= -4 \ln \left(\cosh \frac{\zeta_0}{4}\right) \nabla _{S} N'|_{r=1}\nonumber\\
&& = \frac{3}{2}U' \sin \theta \ {\bf \hat{\theta}}.\nonumber
\end{eqnarray}
To proceed, we eliminate the pressure field from the Stokes equation by evaluating  the curl of this equation. 
This will gives us: 
\begin{equation}\label{CurlStocks}
\nabla\times(\nabla^2 {\bf V'})=\nabla^2(\nabla\times{\bf V'})=0.
\nonumber
\end{equation}
In terms of stream function $H(r,\theta)$, the velocity field can be written as:
\begin{equation}\label{V}
{\bf V'}=\frac{1}{r \sin \theta}\frac{\partial H}{\partial r}\ {\hat {\pmb \theta}} - 
\frac{1}{r^2 \sin \theta}\frac{\partial H}{\partial \theta}\ {\bf \hat{r}},
\nonumber
\end{equation}
where the incompressibility condition $\nabla \cdot {\bf V'}=0$, is assumed. 
According to the boundary condition at infinity, one can suggest a solution for $H$ as:
\begin{equation}
H (r,\theta)= f(r) \sin^2 \theta.
\nonumber
\end{equation}
using the above ansatz, the curl of velocity field reads as:
\begin{eqnarray}\label{Crl V}
\nabla \times \mathbf{V'}&&=\left(\dfrac{1}{r}\dfrac{d^2 f}{d r^2}-\dfrac{2}{r^3}f\right)\sin \theta\ {\hat{\pmb \varphi}}\nonumber\\
&& = A(r) \sin \theta \ {\hat{\pmb \varphi}},
\nonumber
\end{eqnarray}
where $A(r) = \dfrac{1}{r}\dfrac{d^2 f}{d r^2}-\dfrac{2}{r^3}f$. Now we can write:
\begin{eqnarray}
&&\nabla^2 (\nabla\times\mathbf{V'}) 
  = {\hat {\pmb \varphi}}\ \sin \theta \left[\dfrac{d^2 A}{d r^2}+\dfrac{2}{r}\dfrac{d A}{d r}-\dfrac{2A}{r^2}\right]=0.
  \nonumber
\end{eqnarray}
So we have following equation for $A(r)$:
\begin{equation}
\dfrac{d^2 A}{d r^2}+\dfrac{2}{r}\dfrac{d A}{d r}-\dfrac{2A}{r^2}=0,\nonumber
\end{equation}
and its solution is:
\begin{equation}
  A=B'_1 r+\dfrac{B'_2}{r^2},\nonumber
\end{equation}
where $B'_{1}$, $B'_{2}$ are integration constants.
Function $f$ satisfies the following differential equation:
\begin{equation}
  \dfrac{d^2 f}{d r^2}-\dfrac{2}{r^2}f - B'_1 r^2 +\dfrac{B'_2}{r}=0,
  \nonumber
\end{equation}
which has a general  solution like:
\begin{equation}
 f = B_1 r^4 + B_2 r + C_1 r^2 +\dfrac{C_2}{r}.
 \nonumber
\end{equation}
Collecting all the above results, we can write the fluid velocity field as: 
\begin{eqnarray}
&&{\bf V}' = -\frac{2}{r^2}\left(B_1 r^4 + B_2 r + C_1 r^2 +\frac{C_2}{r}\right)\cos \theta \ {\bf \hat{r}}\nonumber\\
&&+\frac{1}{r}\left(4 B_1 r^3 + B_2 + 2 C_1 r - \frac{C_2}{r^2}\right)\sin \theta \ {\hat {\pmb \theta }},
\nonumber
\end{eqnarray}
where  $B_1$, $B_2$, $C_1$ and $C_2$ are constants that can be easily determined by applying the boundary conditions. 
Finally, the fluid velocity field due to a moving Janus particle in a co-moving reference frame  reads as:
\begin{eqnarray}
{\bf V'} &&= -U'\ {\bf \hat{z}} +\frac{1}{2} \frac{U'}{r^3}\left(2 \cos \theta \ {\bf \hat{r}}+ \sin \theta \ {\hat {\pmb {\theta}}}\right) \nonumber\\
&&=- {\bf U'} - \frac{1}{2}{\bf U'} \cdot \left(\frac{\mathbb{I}-3{\bf \hat{r}}{\bf \hat{r}}}{r^3}\right).
\nonumber
\end{eqnarray}

\section{Indirect contribution to the interaction}
\label{appendix:b}
In this appendix we present the calculations that give the indirect contribution to the interaction between two 
Janus particles. We just consider the contribution to the velocity of the first motor, then the symmetry 
arguments will help us 
to write the velocity change of the second motor as well. 
As in the case of a single Janus particle, we will benefit the Lorentz reciprocal theorem to extract the indirect 
contribution.  To use the reciprocal theorem, we need to define the set of two 
hydrodynamic problems which share a common geometry but with different boundary conditions. 
Let's start with defining the problem II, first. The 
case II, corresponds to our main problem defined in the section  "indirect contribution". 
We put ${\bf V}_{\rom{2}}={\bf V}^{\text{ind}}$ and it is subjected to the following boundary conditions:
\begin{equation}
{\bf V}_{\rom{2}}|_{S_1} = {\bf U}_1^0 + {\bf U}_{1}^{\text{ind}} +
{\bf \Omega}_{1}^{\text{ind}}\times {\bf r}_1+ {\bf V}_{S_1}^{\text{ind}},\nonumber
\end{equation}
where we would like to find ${\bf U}_{1}^{\text{ind}}$ and ${\bf \Omega}_{1}^{\text{ind}}$ and 
note that the calculations should be done in laboratory frame.  
Here ${\bf V}_{S_1}^{\text{ind}}$ is the slip velocity given by Equation~\ref{slipind}, and ${\bf U}_1^0$ is the 
first particle velocity from Equation~\ref{barevelocity}.

As we are looking for 6 unknown variables (the components of ${\bf U}_{1}^{\text{ind}}$  and ${\bf \Omega}_{1}^{\text{ind}}$), 
we can consider 6 different  choices for   problem \rom{1}.
To evaluate 3 components for ${\bf U}_{1}^{\text{ind}}$, we choose  ${\bf V}_{\rom{1}}$ as the 
velocity field of a translating particle with velocity $u_{\rom{1}}$ in the three major Cartesian directions, 
and in order to 
evaluate 3 components of ${\bf \Omega}_{1}^{\text{ind}}$, we  choose the velocity field of a rotating particle 
with velocity  $\omega_{\rom{1}}$ in the three major directions. On the surface of the first particle, 
we have ${\bf V}_{\rom{1}}|_{S_1} = {\bf u}_{\rom{1}} + {\pmb \omega}_{\rom{1}} \times {\bf r}_1 $, where:
\begin{eqnarray}\label{differentj}
&&({\bf u}_{\rom{1}},{\pmb \omega}_{\rom{1}})_{j=1}  = (u_{\rom{1}} {\bf \hat{x}},0), ~~~~~  ({\bf u}_{\rom{1}},{\pmb \omega}_{\rom{1}})_{j=4}  = (0,\omega_{\rom{1}} {\bf \hat{x}}),\nonumber\\
 &&({\bf u}_{\rom{1}},{\pmb \omega}_{\rom{1}})_{j=2}  = (u_{\rom{1}} {\bf \hat{y}},0), ~~~~~  ({\bf u}_{\rom{1}},{\pmb \omega}_{\rom{1}})_{j=5}  = (0, \omega_{\rom{1}} {\bf \hat{y}}),\nonumber\\
&&({\bf u}_{\rom{1}},{\pmb \omega}_{\rom{1}})_{j=3}  = (u_{\rom{1}} {\bf \hat{z}},0) , ~~~~~ ({\bf u}_{\rom{1}},{\pmb \omega}_{\rom{1}})_{j=6} = (0, \omega_{\rom{1}} {\bf \hat{z}}),\nonumber
\end{eqnarray}
where $j=1,~2,~3$ denote the cases of translation along the directions 
${\bf \hat{x}}, {\bf \hat{y}}$ and ${\bf \hat{z}}$, respectively, and $j=4,~5,~6$ denote the 
corresponding cases for rotations along ${\bf \hat{x}}, {\bf \hat{y}}$ and ${\bf \hat{z}}$.
Note that for all of the above 6 cases,  on the surface of the second particle we have: ${\bf V}_{\rom{1}}|_{S_2} = 0$.  
Substituting these choices into the left-hand-side of Equation~\ref{Lorentz}, we will obtain:
\begin{eqnarray}
&&\int {\bf V}_{I} \cdot {\pmb \sigma}_{II} \cdot {\bf \hat{n}}\ dS =\int_{S_1} \left({\bf u}_{\rom{1}} + {\pmb \omega}_{\rom{1}} \times {\bf r}_1\right)\cdot {\pmb \sigma}_{\rom{2}} \cdot {\bf \hat{n}}\ dS, \nonumber\\
&& ={\bf u}_{\rom{1}} \cdot \int_{S_1} {\pmb \sigma}_{\rom{2}} \cdot {\bf \hat{n}}\ dS + {\pmb \omega}_{\rom{1}}  \cdot \int_{S_1} {\bf r}_1\times {\pmb \sigma}_{\rom{2}} \cdot {\bf \hat{n}} \ dS, \nonumber\\
&& = {\bf u}_{\rom{1}} \cdot {\bf F}_1 + {\pmb \omega}_{\rom{1}} \cdot {\bf L}_1 = 0,\nonumber
\end{eqnarray}
where ${\bf F}_{\rom{2}}$ and ${\bf L}_{\rom{2}}$ are the force and torque exerted on the particle 1 and are 
equal to zero. Then from the right-hand-side of Equation~\ref{Lorentz} we have:
\begin{eqnarray}
&&\left({\bf U}_{1}^{0} + {\bf U}_{1}^{\text{ind}}\right) \cdot \int_{S_1} {\pmb \sigma}_{\rom{1}} \cdot {\bf \hat{n}}\ dS\ +\  {\bf \Omega}_{1}^{\text{ind}} \cdot \int_{S_1} {\bf r}_1\times {\pmb \sigma}_{\rom{1}} \cdot {\bf \hat{n}} \ dS,\nonumber\\
&& +\ \int_{S_1} {\bf V}_{S_1}^{\text{ind}} \cdot {\pmb \sigma}_{\rom{1}} \cdot {\bf \hat{n}}\ dS = 0,\nonumber
\end{eqnarray}
where ${\bf F}_{\rom{1}}= \int_{S_1} {\pmb \sigma}_{\rom{1}} \cdot {\bf \hat{n}}\ dS$ and 
${\bf L}_{\rom{1}} = \int_{S_1} {\bf r}_1\times {\pmb \sigma}_{\rom{1}} \cdot {\bf \hat{n}} \ dS$ 
are the force and torque exerted on the particle in the problem \rom{1}.
Now for the first three problems ($j=1,~2,~3$) which:
\begin{equation}
{\pmb \omega}_{\rom{1}} = 0, ~~~~ {\bf F}_{\rom{1}} =- 6\pi {\bf u}_{\rom{1}}, ~~~~{\bf L}_{\rom{1}} = 0,~~~~ {\pmb \sigma}_{\rom{1}} \cdot {\bf \hat{n}} =-\frac{3}{2} {\bf u}_{\rom{1}},\nonumber
\end{equation}
we have:
\begin{equation}
-6\pi \left({\bf U}_{1}^{0} + {\bf U}_{1}^{\text{ind}}\right) \cdot {\bf u}_{\rom{1}} +\ \int_{S_1} {\bf V}_{S_1}^{\text{ind}} \cdot {\pmb \sigma}_{\rom{1}} \cdot {\bf \hat{n}}\ dS = 0,\nonumber
\end{equation}
and
\begin{eqnarray}
&&\int_{S_1} {\bf V}_{S_1}^{\text{ind}} \cdot {\pmb \sigma}_{\rom{1}} \cdot {\bf \hat{n}}\ dS = -\frac{3}{2} \left( \int_{S_1} {\bf V}_{S_1}^{\text{ind}} dS\right) \cdot {\bf u}_{\rom{1}},\nonumber\\
&& =-6\pi {\bf U}_1^0 \cdot {\bf u}_{\rom{1}} -6\pi \frac{a_2^3}{D^3} {\bf U}_2^0 \cdot \left(\mathbb{I}-3{\bf \hat{D}}{\bf \hat{D}}\right) \cdot {\bf u}_{\rom{1}}.\nonumber
\end{eqnarray}
Thus the desired velocity due to indirect interaction is obtained as follows:
\begin{equation}
{\bf U}_{1}^{\text{ind}} = \frac{e^3}{D^3}U_2^0\ {\bf \hat{t}}_2 \cdot \left(\mathbb{I}-3{\bf \hat{D}}{\bf \hat{D}}\right).\nonumber
\end{equation}
With similar calculations for the other three problems ($j=3,~4,~5$), 
we can find ${\bf \Omega}_{1}^{\text{ind}}$ as follows:
\begin{equation}
{\bf \Omega}_{1}^{\text{ind}} = -\frac{9}{2} \frac{e^3}{D^4} U_2^0 \left( {\bf \hat{t}}_2\times{\bf \hat{D}}\right).\nonumber
\end{equation}
Similarly, for the second particle we have:
\begin{eqnarray}
&&{\bf U}_{2}^{\text{ind}} = \frac{1}{D^3}U_1^0\ {\bf \hat{t}}_1 \cdot \left(\mathbb{I}-3{\bf \hat{D}}{\bf \hat{D}}\right),\nonumber\\
&&{\bf \Omega}_{2}^{\text{ind}} =  \frac{9}{2} \frac{e}{D^4} U_1^0 \left( {\bf \hat{t}}_1\times{\bf \hat{D}}\right).\nonumber
\end{eqnarray}

\section{Direct contribution to the interaction}
\label{appendix:c}
We have denoted the overall velocity field of the full problem of interacting Janus particles by ${\bf V}$ and, it is constrained to the 
boundary conditions given by Equation~\ref{Velocity Boundary Cond.-dir}. 
Here we apply the reciprocal theorem to extract the direct hydrodynamic interaction between Janus particles. 
In the absence of direct hydrodynamic interaction, we denote the velocity field produced by the 
second Janus particle as: ${\bf u}_\infty({\bf r})$. The first Janus particle is floated in this filed and it is subjected 
to a proper boundary conditions. To use the reciprocal theorem, we consider the case of problem \rom{2} as 
${\bf V}_{\rom{2}}= {\bf V} - {\bf u}_\infty ({\bf r})$ and it is subjected to the following conditions:
\begin{eqnarray}
&&{\bf V}_{\rom{2}}|_{S_1}= {\bf U}_1^0 + {\bf U} _1^{\text{ind}} + {\bf \Omega}_1^{\text{ind}} \times {\bf r}_1 + {\bf V}_{S_1}^{\text{ind}} \nonumber\\
&&~~~~~~~~~+{\bf U} _1^{\text{dir}} + {\bf \Omega}_1^{\text{dir}} \times {\bf r}_1 - {\bf u}_\infty (0)\nonumber\\
&&{\bf V}_{\rom{2}}|_{S_2} = 0.\nonumber
\end{eqnarray}
where $ {\bf u}_\infty (0)$ is the flow field due to the second particle given at the center of the first particle.  
$ {\bf U}_{1}^{\text{ind}}$ and ${\bf \Omega}_{1}^{\text{ind}}$ are the first janus particle velocities due to the indirect 
hydrodynamic interaction and ${\bf V}_{S_1}^{\text{ind}}$ is the slip condition that is given by 
Equation~\ref{slipind}. Velocities ${\bf U} _1^{\text{dir}}$ and ${\bf \Omega}_1^{\text{dir}}$ are unknowns that we 
want to find. So we have six unknown components and we should apply Lorentz theorem six times, as we did in 
appendix~\ref{appendix:b}.
We consider the problem of case \rom{1}, as a spherical particle which moves with constant translational 
and rotational velocities given by: ${\bf u}_{I}$ and ${\pmb \omega}_{I}$. This sphere is immersed in an external velocity field  
given by: 
${\bf u}_\infty ({\bf r})$ and it is subjected to no slip boundary condition.
Therefore, in the laboratory reference 
frame, on the surface of Janus particles we have: 
${\bf V}_{\rom{1}}|_{S_1} = {\bf u}_{\rom{1}} + {\pmb \omega}_{\rom{1}} \times {\bf r}_1 -  {\bf u}_{\infty} (0) $ and  
${\bf V}_{\rom{1}}|_{S_2} =0$. To evaluate the different components of  unknown velocities, we will choose 
six different  choices  for ${\bf u}_{I}$ and ${\pmb \omega}_{I}$ as have been chosen in appendix~B.  

Substituting into Equation~\ref{Lorentz} then, simplifying the results, we will arrive at:
\begin{eqnarray}
&&\left({\bf U}_1^0 +{\bf U}_1^{\text{dir}} + {\bf U}_1^{\text{ind}}\right) \cdot {\bf F}_{\rom{1}} + \left({\bf \Omega}_1^{\text{dir}}  + {\bf \Omega}_1^{\text{ind}}\right) \cdot {\bf L}_{\rom{1}}\nonumber\\
&& = \int_{S_1} {\bf u}_\infty \cdot {\pmb \sigma}_{\rom{1}} \cdot {\bf \hat{n}}\ dS - \int_{S_1} {\bf V}_{S_1}^{\text{ind}} \cdot {\pmb \sigma}_{\rom{1}} \cdot {\bf \hat{n}}\ dS.\nonumber 
\end{eqnarray} 
Now we consider the three problems $j=1,~2,~3$ (the problems that have been defined in appendix~B). 
So the right-hand-side terms are calculated as:
\begin{eqnarray}
&&\int_{S_1} {\bf u}_\infty \cdot {\pmb \sigma}_{\text{\rom{1}}} \cdot {\bf \hat{n}}\ dS = 
-\frac{3}{2} {\bf u}_{\text{\rom{1}}} \cdot \int_{S_1} {\bf u}_\infty\ dS \nonumber\\
&&\approx -\frac{3}{2} {\bf u}_{\text{\rom{1}}} \cdot 4\pi \left({\bf u}_\infty({\bf r}_{1}=0) + \frac{1}{6}\nabla^2{\bf u}_\infty({\bf r}_{1}=0)\right) \nonumber\\
&&\approx -6\pi {\bf u}_{\text{\rom{1}}} \cdot \left(-\frac{1}{2}\frac{e^3}{D^3}\left({\bf U}_2^0+{\bf U}_2^{\text{ind}}\right)\cdot(\mathbb{I}-3 {\bf \hat{ D}}{\bf \hat{ D}}) +  {\cal O} \left(\frac{1}{D^6}\right)\right),\nonumber
\end{eqnarray}
and
\begin{eqnarray}
&&\int {\bf V}_{S_1}^{\text{ind}} \cdot {\pmb \sigma}_{\text{\rom{1}}} \cdot {\bf \hat{n}}\ dS \nonumber\\
&&=-\frac{9}{4}\left({\bf U}_{1}^{0}+{\bf U}_{1}^{\text{ind}}\right)\cdot \int _{S_1} ({\bf \hat{ r}}{\bf \hat{ r}}-\mathbb{I}) \ d\cos\theta \ d\varphi \cdot {\bf u}_{\text{\rom{1}}}\nonumber\\
&&= 6\pi \left({\bf U}_{1}^{0}+{\bf U}_{1}^{\text{ind}}\right)\cdot {\bf u}_{\text{\rom{1}}},\nonumber
\end{eqnarray}
now collecting the above results, we have:
\begin{eqnarray}
&&-6\pi \left({\bf U}_1^0 +{\bf U}_1^{\text{dir}} + {\bf U}_1^{\text{ind}}\right) \cdot {\bf u}_{\text{\rom{1}}}=\nonumber\\
&&  -6\pi {\bf U}_1^0 \cdot {\bf u}_{\text{\rom{1}}} -  6\pi {\bf U}_{1}^{\text{ind}}\cdot {\bf u}_{\text{\rom{1}}} \nonumber\\
&&- 6\pi  \left(-\frac{1}{2}\frac{e^3}{D^3}\left({\bf U}_2^0+{\bf U}_2^{\text{ind}}\right)\cdot(\mathbb{I}-3 {\bf \hat{ D}}{\bf \hat{ D}}) + {\cal O}\left(\frac{1}{D^6}\right)\right) \cdot {\bf u}_{\text{\rom{1}}}.\nonumber
\end{eqnarray}
The final result for  ${\bf U}_{1}^{\text{dir}}$ can be written as:
\begin{eqnarray}
{\bf U}_1^{\text{dir}}&&= -\frac{1}{2}\frac{e^3}{D^3}{\bf U}_2^0\cdot(\mathbb{I}-3 {\bf \hat{ D}}{\bf \hat{ D}}) 
-\frac{1}{2}\frac{e^3}{D^3}{\bf U}_2^{\text{ind}}\cdot(\mathbb{I}-3 {\bf \hat{ D}}{\bf \hat{ D}})\nonumber\\
&& +{\cal O} \left(\frac{1}{D^6}\right).\nonumber
\end{eqnarray}
Now for evaluating ${\bf \Omega^{\text{dir}}}$, we consider three problems given as  $j=4,~5,~6$ in appendix~B. 
For these cases we have:
\begin{equation}
{\bf u}_{\rom{1}} =0, ~~~~ {\bf F}_{\rom{1}} = 0, ~~~~{\bf L}_{\rom{1}} = -8\pi {\pmb \omega}_{\rom{1}},~~~~ {\pmb \sigma}_{\rom{1}} \cdot {\bf \hat{n}} =-3\ {\pmb \omega}_{\rom{1}} \times \hat{{\bf r}}_{1}.\nonumber
\end{equation}
So the right-hand-side terms of the reciprocal integral are calculated as:
\begin{eqnarray}
&&\int_{S_1} {\bf u}_\infty \cdot {\pmb \sigma}_{\text{\rom{1}}} \cdot {\bf \hat{n}}\ dS \nonumber\\
&&= -3\ {\pmb \omega}_{\text{\rom{1}}} \cdot \int_{S_1} {\bf \hat{r}} \times {\bf u}_\infty\ dS \nonumber\\
&&\approx -3\ {\pmb \omega}_{\text{\rom{1}}} \cdot \int_{S_1} {\bf \hat{r}} \times \left({\bf u}_\infty({\bf r}_{1}=0) + \frac{4 \pi}{3}\nabla \times {\bf u}_\infty({\bf r}_{1}=0)\right) \nonumber\\
&&\approx {\cal O} \left(\frac{1}{D^9}\right),\nonumber
\end{eqnarray}
and
\begin{eqnarray}
&&\int {\bf V}_{S_1}^{\text{ind}} \cdot {\pmb \sigma}_{\text{\rom{1}}} \cdot {\bf \hat{n}}\ dS =-3 \ {\pmb \omega}_{\text{\rom{1}}} \cdot \int _{S_1} {\bf \hat{ r}}\times {\bf V}_{S_1}^{\text{ind}}\ d\cos\theta \ d\varphi \nonumber\\
&&= - 36\pi \frac{e^{3}}{D^{4}} \left({\bf U}_{2}^{0} \times {\bf \hat{D}}\right) \cdot {\pmb \omega}_{\text{\rom{1}}},\nonumber\\
&&= 8 \pi\ {\bf \Omega}^{\text{ind}} \cdot {\pmb \omega}_{\text{\rom{1}}}.\nonumber
\end{eqnarray}
Substituting the above results into the reciprocal integral, we can see that:
\begin{equation}
-8\pi \left({\bf \Omega}^{\text{ind}} + {\bf \Omega}^{\text{dir}} \right) \cdot {\pmb \omega}_{\text{\rom{1}}}= - 8 \pi\ {\bf \Omega}^{\text{ind}} \cdot {\pmb \omega}_{\text{\rom{1}}} + {\cal O} \left(\frac{1}{D^9}\right).\nonumber
\end{equation}
This shows that  the  rotational velocity has no direct contribution behaving stronger than a term like $(1/D)^9$.

\bibliographystyle{unsrt}
\bibliography{2sElectro}

\end{document}